\documentclass{emulateapj}

\usepackage{hyperref}

\begin{document}


\title{An Orphan No Longer? Detection of the Southern Orphan Stream and a Candidate Progenitor}

\author{Carl J. Grillmair}
\affil{Infrared Processing and Analysis Center, California Institute of Technology, Pasadena, CA 91125}
\email{carl@ipac.caltech.edu}

\author{Lauren Hetherington, Raymond G. Carlberg}
\affil{Department of Astronomy and Astrophysics, University of Toronto, Toronto, ON M5S 3H4, Canada}
\email{hetherington@astro.utoronto.ca}
\email{carlberg@astro.utoronto.ca}

\author{Beth Willman}
\affil{Haverford College, Department of Physics and Astronomy, Haverford, PA 19041}
\email{beth.willman@gmail.com}

\begin{abstract}

Using a shallow, two-color survey carried out with the Dark Energy
Camera, we detect the southern, possibly trailing arm of the Orphan
Stream. The stream is reliably detected to a declination of
-38\arcdeg, bringing the total known length of the Orphan stream to
$108\arcdeg$. We find a slight offset or ``S'' shape in the stream at
$\delta \simeq -14\arcdeg$ that would be consistent with the
transition from leading to trailing arms.  This coincides with a
moderate concentration of $137 \pm 25$ stars (to $g = 21.6$) that we
consider a possible remnant of the Orphan progenitor. The position of
this feature is in agreement with previous predictions.

\end{abstract}



\section{Introduction}

The Orphan stream was among the first stellar debris streams detected
in the Sloan Digital Sky Survey (SDSS) \citep{belokurov2006,
  grillmair2006c, belokurov2007}. Populous and roughly $2\arcdeg$ wide
on the sky, the stream is clearly much broader and stronger than known
globular streams such as Pal 5
\citep{odenkirchen2003,grillmair2006a}. This, along with subsequent
findings of a metallicity dispersion of $\sigma$[Fe/H] = 0.56 dex
\citep{casey2013} and a metallicity gradient amplitude of 0.3 dex
\citep{sesar2013} led researchers to conclude that the Orphan stream
must be the remnant of a dwarf galaxy. Early modeling efforts
suggested that the stream might be related to the neutral hydrogen
Complex A \citep{fellhauer2007, jin2007}, and that the progenitor of
the stream might be the nearby dwarf galaxy UMa II
\citep{fellhauer2007}. However, subsequent work by \citet{sales2008}
and \citet{newberg2010} does not support these ideas.

\citet{newberg2010} used measured positions and velocities to derive
an orbit of the stream, and determined that the orbit is prograde,
moderately inclined to the Galactic plane ($i \approx 34\arcdeg$),
fairly eccentric ($e \approx 0.7$), extending out to $\approx 90$ kpc
from the Galactic center, and that the portion of the stream visible
in the SDSS footprint is the leading arm. Based on the rising surface
density of the stream at the southern edge of the SDSS footprint (in
the direction of decreasing Galactocentric radius and far from
apogalacticon), they also suggested that the progenitor would most
likely be found between declinations of $0\arcdeg$ and $-16\arcdeg$.

In this letter we describe the first results of a shallow imaging survey
designed to trace the Orphan Stream well south of the SDSS footprint.
We briefly describe the observations in Section
\ref{observations}. We analyze the spatial and color-magnitude
characteristics of the stream in Section \ref{analysis}.  Concluding
remarks are given in Section \ref{conclusions}.

\section{Observations} \label{observations}

Using the Orphan orbit estimation of \citet{newberg2010} as a guide,
we imaged a $9\arcdeg$ to 15\arcdeg-wide swath of sky extending from
the celestial equator to $\delta \simeq -53\arcdeg$ and covering an
area of 487 deg$^2$. This was carried out during just two observing
nights using the remarkably efficient Dark Energy Camera (DECam) on
the Blanco 4-meter telescope at the Cerro Tololo Interamerican
Observatory (CTIO). Observations were made in $g$ and $i$, and
exposures were kept to two 30 second dithers per field to maximize the
area covered while still reaching well past the main sequence turnoff
of the stream.  Observations were carried out over two observing
seasons, with one night in March of 2014 and another in March of
2015. Conditions were photometric during both nights, with typical
seeing of $0.9\arcsec$ in $i$ and $1-1.2\arcsec$ seeing in $g$, though
with excursions of $> 2\arcsec$ for a short period during the
2014 run.

The resulting 6.3 TB of data were processed using the 2015 version of
the DECam Community Pipeline \citep{valdes2014}. (2014 data were
reprocessed with the 2015 pipeline to take advantage of several
improvements) The data were subsequently transferred to the University
of Toronto, where a photometry pipeline based on SExtractor and PSFEx
\citep{bertin1996} was constructed to photometer individual images
using point spread function (PSF) fitting.

PSFs, aperture corrections, and 2nd order color terms were computed
for each individual detector. The photometry was calibrated against
the SDSS catalog using $\simeq 20$ deg$^2$ of imaging in the Sloan
footprint. Average atmospheric extinction coefficients for CTIO were
used throughout. Stars were typically observed at least twice in each
filter (with the exception of a small number of stars falling within
the CCD gaps), and the individual photometric measurements were
combined over all relevant fields and over both observing runs.
Within the SDSS footprint, calibration is good to 0.02 mags
RMS.

Perhaps owing to the variable nature of the PSFs over a field as large
as that of DECam, we found that the star/galaxy separation parameter
``CLASS\_STAR'' was rather unreliable, with a spread that varied
considerably from the center to the edge of each field. Hence we
relied primarily on the ``FWHM\_WORLD'' and ``ELLIPTICITY'' parameters
to excise sources that were clearly extended. Imposing limits of
FWHM\_WORLD $ < 3\arcsec$, ELLIPTICITY $< 0.2$, FLAGS=0, and $16 < g <
21.6$ reduced a catalog of 15 million sources to 3.5 million. The
FWHM\_WORLD and ELLIPTICITY cuts were deliberately somewhat generous,
as tighter constraints resulted in an obvious diminution of source
counts from the center to the edge of each field. These limits
necessarily entail the inclusion of some background galaxies, which
will contribute additional noise to the filtered maps, but with a
limit of $g = 21.6$ this should not be excessive.

Some calibration issues remain unresolved. For example, star counts
appear rather more sensitive to airmass than we expect. While many of
our fields are essentially complete to $g \approx 23$, others (with
airmasses $\ge 2$) are complete to only $g \approx 21.7$. These issues
will be further explored in a forthcoming contribution. For our present
purposes, we avoid these issues by simply cutting off our sample at
$g = 21.6$.

\section{Analysis} \label{analysis}

We used a matched filter to optimally separate the metal poor stars of
the Orphan Stream from the much larger population of foreground disk
stars \citep{rock2002, grillmair2009}. This technique has been used to
detect several streams at surface densities as low as 10 stars
deg$^{-2}$
\citep{grillmair2006a,grillmair2006b,grillmair2006c,grillmair2009,grillmair2011,bonaca2012}. We
generated a filter based on the Padova database of theoretical stellar
isochrones \citep{marigo2008, girardi2010}, selecting for stars with
[Fe/H] = -1.6.  All stars with $16 < g < 21.6$ were used, and we
dereddened the photometry as a function of position on the sky using
the DIRBE/IRAS dust maps of \citet{schleg98}, corrected using the
prescription of \citet{schlafly2011}. The foreground population was
sampled in stream-free regions extending along the edges of our survey
area. Figure 1 shows the filtered star count distribution using a filter
based on an isochrone with Z = 0.0005 and an age of 12 Gyrs, optimized
for populations at a distance of 18 kpc. 

Nearly centered within the survey area is a long, broad feature
extending to nearly $-40\arcdeg$. The 18 kpc
distance used in Figure 1 corresponds to the strongest stream signal
and roughly matches the 19-21 kpc range of distances expected on the
basis of an orbit fit to Newberg et al. (2010)'s data compilation for
the northern Orphan stream. Differences may be due to inaccurate
matching of the DECam $g$ and $i$ photometry to the Sloan filters
assumed by the Padova isochrones, or possibly a metallicity gradient
in the Orphan Stream \citep{sesar2013}. It may also be that 18 kpc is
the correct distance of the stream in this region, and that the actual
orbit of stream stars in this region needs to be refined.

The northern $10\arcdeg$ of the detected stream matches nicely with
the portion of the stream detected in the Sloan footprint.  A
full-width-at-half-maximum of $\approx 1.5-2\arcdeg$ is also
consistent with that observed in the northern stream. The stream
appears to be reliably detected to $\delta \simeq -38\arcdeg$, below
which the character of the distribution changes significantly (see
below).  This brings the known length of the stream to $\approx
108\arcdeg$. Over the southern interval $-18\arcdeg < \delta <
-38\arcdeg$, we find the stream is well fit (to within $0.25\arcdeg$)
by a polynomial of the form:

\begin{equation}
\label{trace}
\alpha = 163.147 - 0.0896 \times \delta + 0.00804 \times \delta^2 
\end{equation}

Figure 2 shows the distribution of $E(B-V)$ over our survey area from
the maps of \citet{schleg98}. A comparison of Figures 1 and 2 shows
that the pattern of the star counts in the region $-39\arcdeg > \delta
> -45\arcdeg$ closely matches the filamentary distribution of dust
emission and enhanced reddening. Dereddening our photometry has
evidently pulled an excess of fainter and redder stars into the
sample. Whereas $E(B-V)$ is fairly uniform and ranges from 0.02 to
0.06 over the northern half of the survey area, the filamentary
structures at $\delta \approx -43\arcdeg$ show color excesses ranging
from 0.2 to 0.3. Arbitrarily scaling down the \citet{schlafly2011}
absorption coefficients reduces the effect, but does not yield any
convincing signatures of an underlying stream. Tracing the stream
through this region would presumably benefit in the near term from a deep,
near-infrared survey, though it should ultimately be detected
in Gaia proper motion data.

Figure 3 shows a color-magnitude distribution of stars chosen to lie
within the $\pm 1\arcdeg$ of the center of the stream north of $\delta
= -36\arcdeg$. Over plotted are isochrones for populations with Z = 0.0001
([Fe/H] = -2.1) and Z=0.0005 ([Fe/H] = -1.6). Z=0.0005 appears to
match the main sequence somewhat better than Z=0.0001, which
corresponds to the metallicity found by \citet{newberg2010} for the blue
horizontal branch stars. The value [Fe/H] = -1.6 used in Figure 1
matches a measurement of [Fe/H] = -1.63 found by
\citet{casey2013} for red giants. Note that \citet{sesar2013} see evidence for a
metallicity gradient in the northern stream, with the nearer, more
southerly stars being $\simeq 0.3$ dex more metal rich than the more
northerly, more distant stars.

Figure 4 shows the southern Orphan stream in greater detail. Over
plotted is an orbit fit to the data collected by \citet{newberg2010}
for the northern Orphan stream. This orbit was computed using the
Galactic model of \citet{allen1991}, which assumes a spherical
halo. The orbit generally matches the trajectory of the southern
stream, though offset somewhat towards the east below $\delta \approx
-14\arcdeg$. There are a number of possible reasons for the offset:
$(i)$ the orbit calculation did not take into account the southern
stream (which as yet has no velocity information), $(ii)$ the effects
of halo flattening or triaxiality have not been considered, or $(iii)$
we may be looking at the trailing arm of the stream.

If we use the northern orbit fit as a guide, we see that while it
appears to fit the stream reasonably well north of $\delta \approx
-14\arcdeg$, an eastward offset of $\approx 1.5\arcdeg$ begins rather
suddenly south of $\delta = -14\arcdeg$, and stays roughly constant to
$\delta = -38\arcdeg$. At (R.A., dec) = (167\arcdeg, -14\arcdeg),
midway between the northern orbit fit and the run of Equation 1, there
is a moderate but significant, $1.5\arcdeg$-wide overdensity of stars
that is somewhat larger and stronger than the clumps to the immediate
north or south. This clump appears to be the extended, northern
portion of a feature found by \citet{newberg2010} in an ``outrigger''
SEGUE stripe at $\delta \approx -15\arcdeg$. We hypothesize that the
transition from the northern to the southern portions of the stream is
the ``S-shape'' signature expected from a progenitor losing stars from
its first and second Lagrange points.  We further suggest that this clump of
stars could be the remnant of the progenitor of the Orphan stream.

Based on the rise and fall of stream surface density with position
along the stream, \citet{newberg2010} predicted that the progenitor of
the Orphan stream should be situated between $\delta = 0\arcdeg$ and
$\delta = -16\arcdeg$. This is consistent with the position of our
overdensity at $\delta \approx -14\arcdeg$. Moreover,
\citet{newberg2010} determined that the northern portion of the Orphan
stream must be the leading arm.  Tidal stripping in a constant-$v_c$
potential requires that the leading arm should be made up of stars
released from progenitor's first Lagrange point, into orbits of lower
Galactocentric radius $R$. Conversely, the trailing arm will be made
up of stars lost from the second Lagrange point, falling behind the
progenitor and orbiting at larger $R$. This is consistent with Figure
4; the westward offset of the southern portion of the stream takes it
further away from the Galactic center, which is to the left in the
figure.

At a distance of 18 kpc, a $1.5\arcdeg$ offset corresponds to $\approx
470$ pc. The L1 and L2 lagrange points will always be aligned along a
radial to the Galactic center. At the current position of the putative
progenitor, we would be viewing it at an angle of $\approx 23\arcdeg$
from the L1 - L2 radial. If indeed the northern and
southern Orphan streams are leading and trailing arms, respectively,
then the implied physical separation would be 1.2 kpc. We consequently
take the upper limit on the tidal radius of the progenitor to be 600
pc.

The number of stars within the putative progenitor is not
large. Examining a square region $1.1\arcdeg$ on a side and centered
on (R.A., dec) = (167.125\arcdeg,-14.273\arcdeg) and comparing with
background fields to the east and west, we count stars with $0.16 < g
- i < 0.44$ and $19.9 < g < 21.6$. Scaling by the area ratios, we find
a background-subtracted count of $137 \pm 24$ stars.  Integrating over
the luminosity function of Omega Cen \citep{demarchi1999}, we arrive
at an approximate total population of $2100 \pm 400$ stars. If indeed
this clump is the progenitor of the Orphan stream, then it would
appear to be virtually the last remnant of the original satellite. The
surface density of the object is proportional to $r^{-(0.7 \pm 0.3)}$,
making it unlikely that the feature could be gravitationally bound.

By definition, the tidal radius $r_t^3 = (M_p/2M_G(R)) R^3$ in a flat
rotation curve, where
$M_G(R)$ is the mass of the Galaxy within Galactocentric radius $R$, and
$M_p$ is the mass of the progenitor. If we take $R = 21$ kpc and
$M_G(R) = 1 - 2 \times 10^{11}$ M$_{\odot}$, we arrive at an upper
limit on the progenitor's recent mass of $\sim 4.7 - 9.3 \times
10^6$ M$_{\odot}$. Depending on the number of red giants, the luminosity of the
object could range from $1\times 10^4$ to $4 \times 10^4$
L$_{\odot}$. If a bound object remains, then $M/L \sim 120 - 930
M_{\odot}/L_{\odot}$.

Using the luminosity-metallicity relation of \citet{kirby2011}, the
[Fe/H] = -1.6 measurement of \citet{casey2013} suggests a total
luminosity of the original progenitor of $2.5 \times 10^6
L_{\odot}$. On the other hand, Newberg et al. (2010)'s value of [Fe/H]
= -2.1 implies $6 \times 10^5 L_{\odot}$.  Our luminosity estimate
above would suggest that the progenitor has lost between 94\% and 100\%
of its original mass. 

There are other surface density peaks evident in Figure 4
but we are less inclined to consider these as progenitor
candidates as they do not show the morphological indicators
(e.g. offsets) we would associate with the transition from leading to
trailing arms. Given the orientation of the Orphan stream and our view
of it, such a feature should be readily apparent.

We note also that near the southernmost end of the survey area is the
globular cluster Ruprecht 106. This cluster is situated along the
plausible extension of the Orphan stream. However, while its
metallicity of [Fe/H] = -1.67 \citep{harris1996} is similar to that of
the Orphan stream, its distance of 12 kpc and radial velocity of -44
km s$^{-1}$ are at odds with values of 21 kpc and +72 km s$^{-1}$
predicted by the orbit fit to the northern Orphan stream. We conclude
that Ruprecht 106 is unlikely to be physically associated with the
stream.

\section{Conclusions} \label{conclusions}

Using a large, shallow DECam survey, we have traced the Orphan stream
from the celestial equator to $\delta \simeq -38\arcdeg$. The stream
appears to be roughly 18 kpc distant, and its trajectory generally agrees with
expectations based on orbit fits to the northern stream. The color magnitude
distribution is clearly metal poor and appears similar to that of the
northern Orphan stream. We find a stellar concentration and apparent offsets in
the stream that would be consistent with a remnant progenitor. 

This southern extension of the Orphan stream should enable significant
improvements in constraining the overall orbit, and ultimately the
shape of the Galactic potential. This is particularly interesting in
that the Orphan stream passes through quadrants of the halo not probed
by the Sagittarius stream. Slightly deeper than the present survey,
the Pan-STARRS survey may enable us to improve the signal-to-noise
ratio somewhat for $\delta > -30\arcdeg$. For more southerly regions,
where we are strongly affected by reddening, a deep,
near-infrared survey may help to trace the stream still further south.

\acknowledgments

We gratefully acknowledge Jonathan Hargis for helpful suggestions in
the course of developing our photometry pipeline.  This project used
data obtained with the Dark Energy Camera (DECam), which was
constructed by the Dark Energy Survey (DES) collaboration. Funding for
the DES Projects has been provided by the DOE and NSF(USA),
MISE(Spain), STFC(UK), HEFCE(UK). NCSA(UIUC), KICP(U. Chicago),
CCAPP(Ohio State), MIFPA(Texas A\&M), CNPQ, FAPERJ, FINEP (Brazil),
MINECO(Spain), DFG(Germany) and the collaborating institutions in the
Dark Energy Survey, which are Argonne Lab, UC Santa Cruz, University
of Cambridge, CIEMAT-Madrid, University of Chicago, University College
London, DES-Brazil Consortium, University of Edinburgh, ETH Zurich,
Fermilab, University of Illinois, ICE (IEEC-CSIC), IFAE Barcelona,
Lawrence Berkeley Lab, LMU Munchen and the associated Excellence
Cluster Universe, University of Michigan, NOAO, University of
Nottingham, Ohio State University, University of Pennsylvania,
University of Portsmouth, SLAC National Lab, Stanford University,
University of Sussex, and Texas A\&M University.

ms
{\it Facilities:} \facility{CTIO:Blanco (DECam)}.

\clearpage

\begin{figure}
\epsscale{1.0}
\plotone{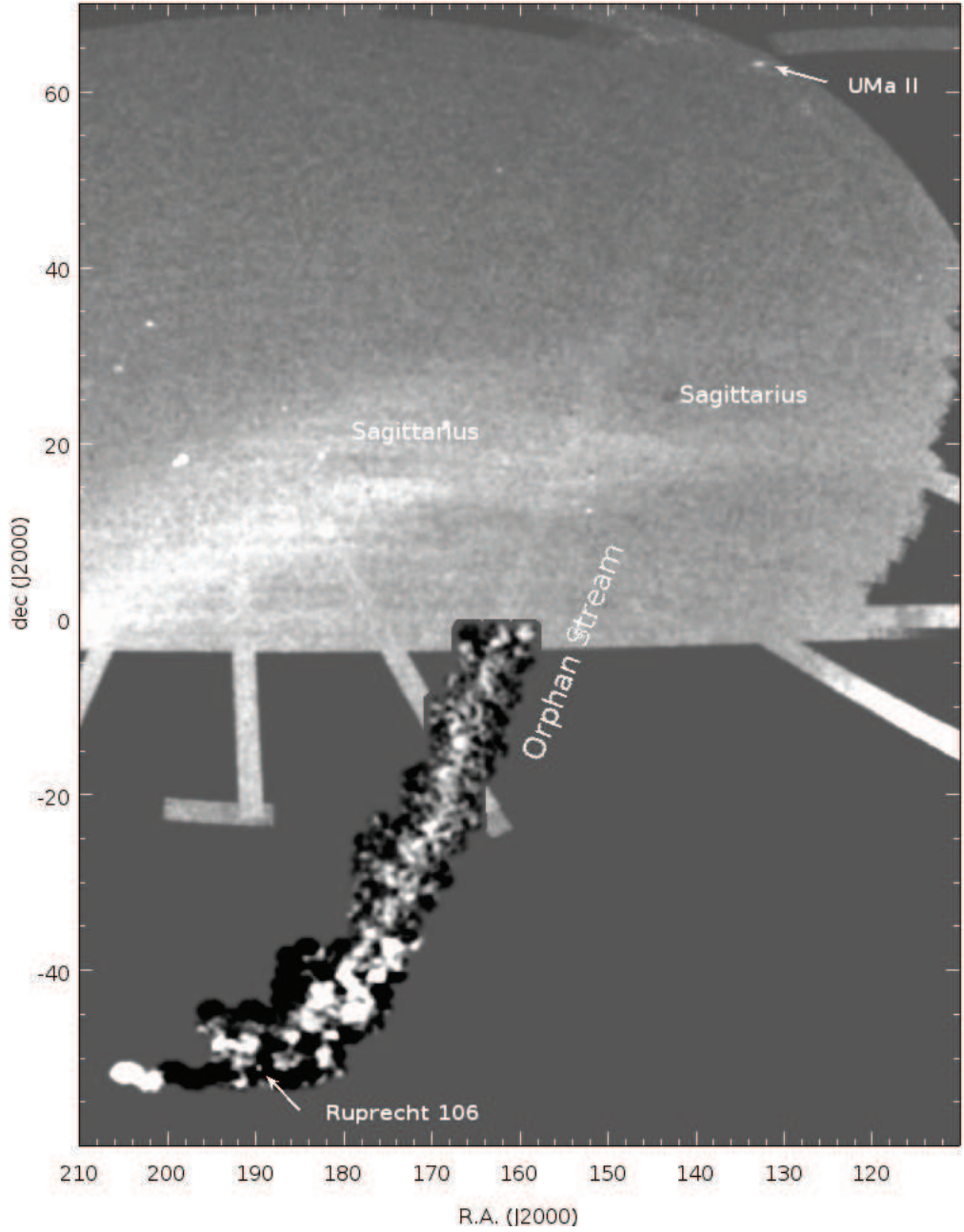}
\caption{Filtered surface density map of our Decam survey area,
  overlaid on the SDSS DR10 footprint.  The stretch is linear, with
  lighter areas indicating higher surface densities. The map is the
  result of a filter based on a Padova isochrone with [Fe/H] = -1.6,
  an age of 12 Gyr, and shifted to a distance of 18 kpc. The Sloan
  data have been smoothed with a $0.3\arcdeg$ Gaussian kernel while
  the DECam map, owing to its somewhat shallower depth, has been
  smoothed with a $0.5\arcdeg$ kernel. Seeing was $0.9-1.8\arcsec$ over most of the survey area, with two stripes ($-37\arcdeg > \delta > -41\arcdeg$, $-45\arcdeg > \delta > -48\arcdeg$) having seeing in excess of $2\arcsec$. The highest airmasses ($> 1.8$) occurred at $\delta > -7\arcdeg$.}

\end{figure}

\begin{figure}
\epsscale{1.0}
\plotone{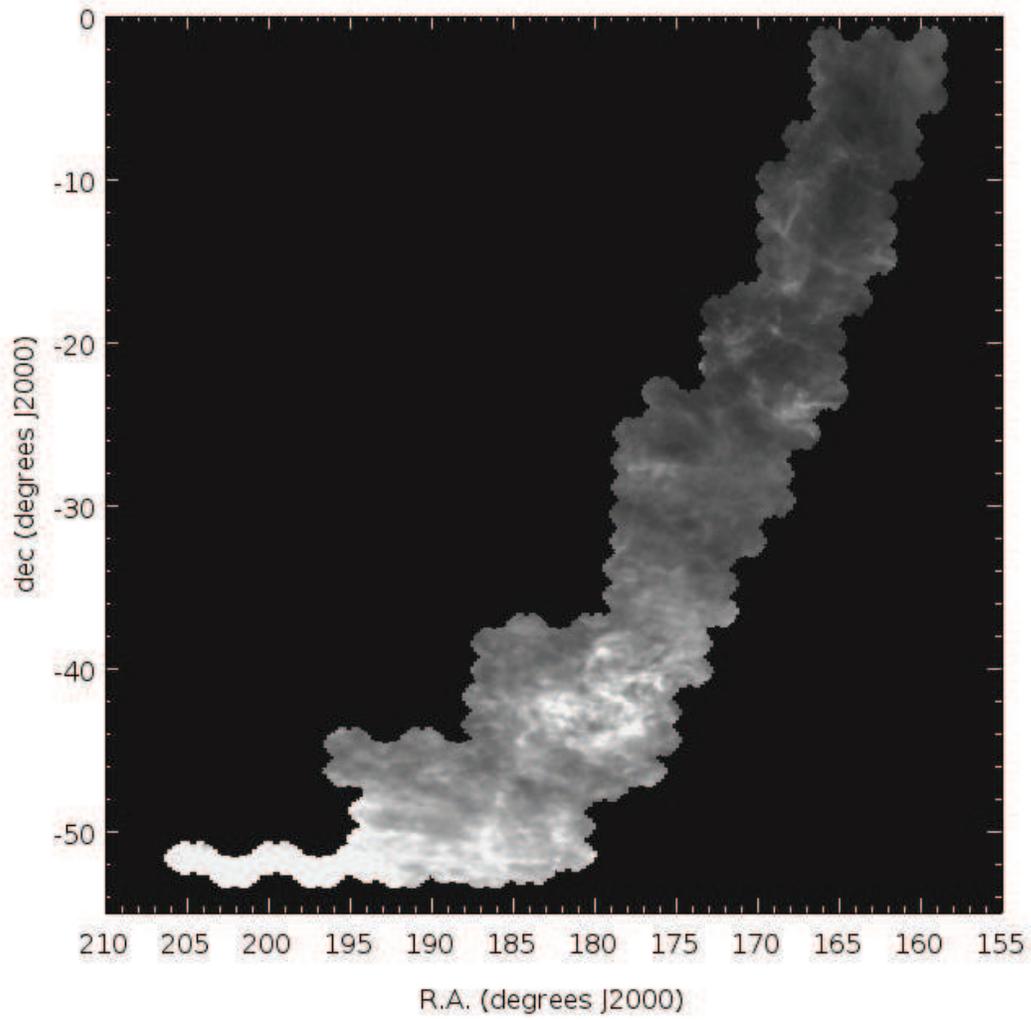}
\caption{Distribution of $E(B-V)$ over the field shown in Figure 1. Lighter areas indicate higher color excesses. Values of the color excess range from 0.02 in the darkest, northern reaches of the survey, to 0.3 in the brightest filaments at $\delta \approx -43\arcdeg$. }

\end{figure}

\begin{figure}
\epsscale{1.0}
\plotone{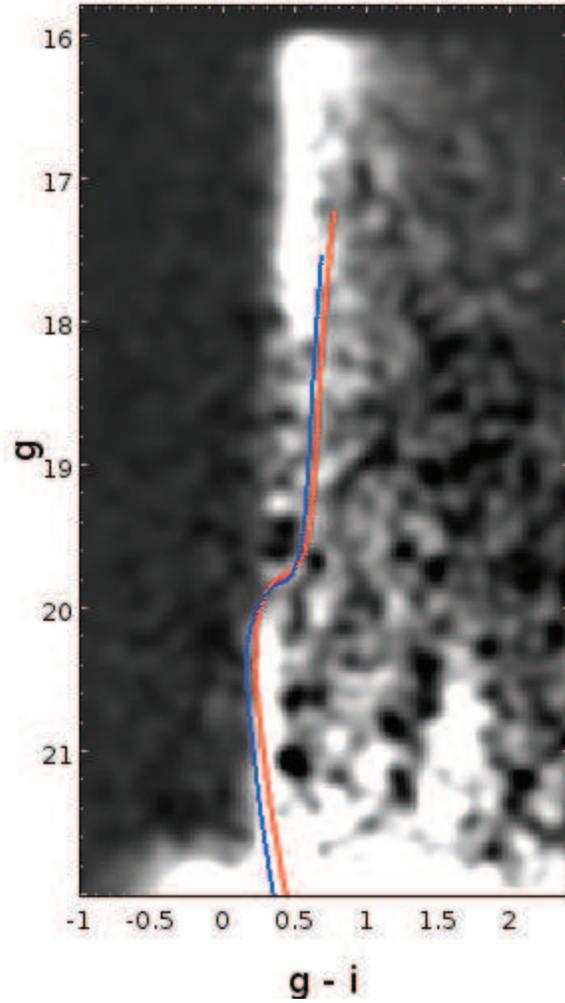}
\caption{Hess diagram of stars lying within $1\arcdeg$ of the
  centerline of the Orphan stream north of $\delta = -36\arcdeg$, after subtraction of the distribution of stars along the edge of the survey area. The result has been convolved with a 0.05 mag Gaussian kernel. 
  Lighter areas indicate higher surface densities.The solid lines
  shows a Padova isochrone with [Fe/H] =-2.1 (blue, left) and -1.6
  (red, right), age 12 Gyrs, and shifted to distances of 20 and 18
  kpc, respectively. }

\end{figure}

\begin{figure}
\epsscale{1.0}
\plotone{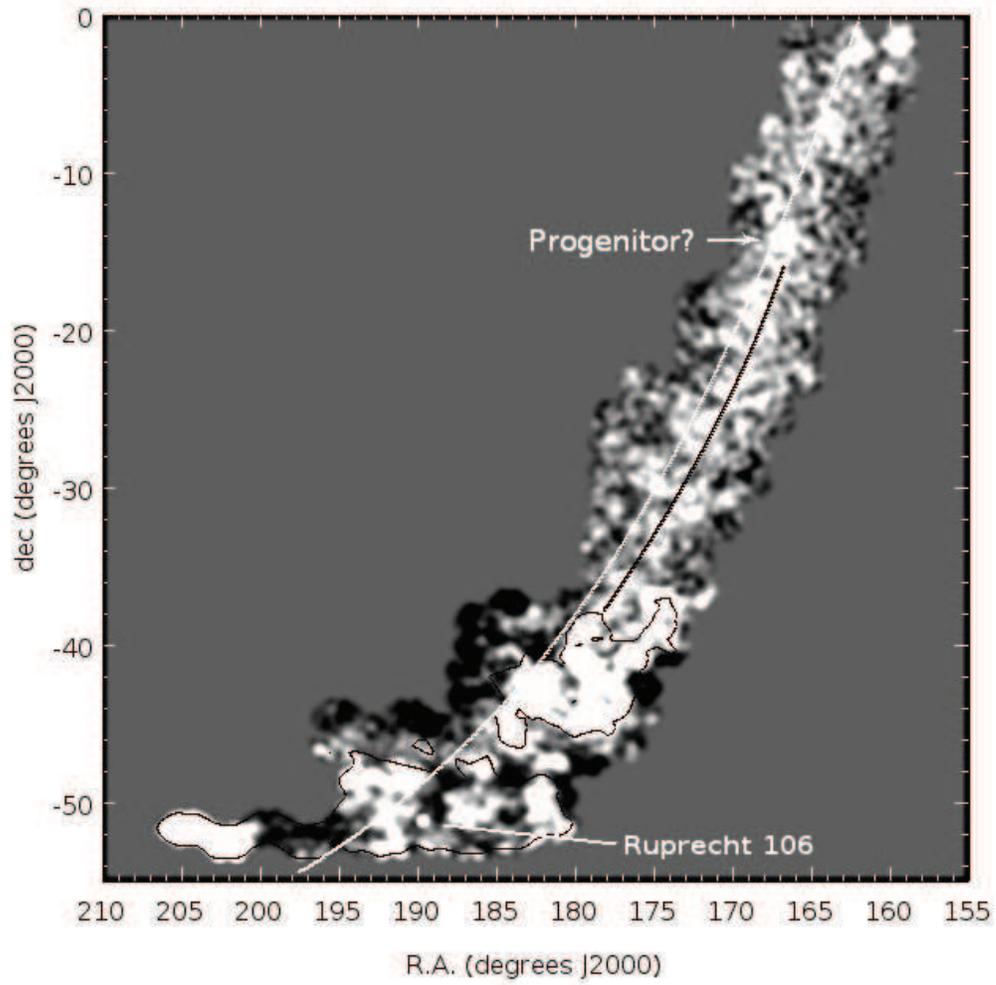}
\caption{A more detailed view of the DECam filtered surface density map in Figure 1, again 
  smoothed with a Gaussian kernel of width $0.5\arcdeg$. The white
  curve traces an orbit fit to the position and velocity data of
  \citet{newberg2010} for the northern Orphan stream. The black curve
  extending to $\delta = -38\arcdeg$ is the fit to the southern
  portion of the stream given by Equation 1. The black contour in the
  southern half of the survey area corresponds to $E(B-V) = 0.12$
  in a smoothed version of Figure 2. Our progenitor
  candidate is indicated. }

\end{figure}

\end{document}